\documentclass[conference]{IEEEtran}
\IEEEoverridecommandlockouts

\usepackage{physics}
\usepackage{cite}
\usepackage{amsmath,amssymb,amsfonts}
\usepackage{algorithmic}
\usepackage{graphicx}
\usepackage{textcomp}
\usepackage{xcolor}

 \usepackage[moderate,tracking=normal]{savetrees}
 
\usepackage{pgffor}
\usepackage{subfigure}
\usepackage{tikz}
\usetikzlibrary{fadings}
\usetikzlibrary{patterns}
\usetikzlibrary{shadows.blur}
\usetikzlibrary{quantikz}
\usepackage{adjustbox}

\usepackage[numbers,compress]{natbib}

\definecolor{quantumweek-1}{RGB}{182,214,186}
\definecolor{quantumweek-2}{RGB}{214,205,182}
\definecolor{quantumweek-3}{RGB}{182,190,214}
\definecolor{quantumweek-4}{RGB}{214,182,192}
\definecolor{quantumweek-5}{RGB}{81,93,130}
\definecolor{quantumweek-6}{RGB}{39,87,45}

\def\BibTeX{{\rm B\kern-.05em{\sc i\kern-.025em b}\kern-.08em
    T\kern-.1667em\lower.7ex\hbox{E}\kern-.125emX}}
\begin{document}

\title{Distributed Quantum Computing for Chemical Applications \\
\thanks{We would like to acknowledge the Government of Canada’s New Frontiers in Research Fund (NFRF), for grant NFRFE-2022-00226, and the Quantum Software Consortium (QSC), financed under grant \#ALLRP587590-23 from the National Sciences and Engineering Research Council of Canada (NSERC) Alliance Consortia Quantum Grants.
}
}

\author{\IEEEauthorblockN{Grier M. Jones\IEEEauthorrefmark{1}\IEEEauthorrefmark{2}\IEEEauthorrefmark{3},Hans-Arno Jacobsen\IEEEauthorrefmark{2}}
\IEEEauthorblockA{
Department of Chemical and Physical Sciences\IEEEauthorrefmark{1}\\
The Edward S. Rogers Sr. Department of Electrical and Computer Engineering\IEEEauthorrefmark{2}\\
University of Toronto\\
Toronto, Ontario, Canada\\
Email: grier.jones@utoronto.ca\IEEEauthorrefmark{3}}

}

\IEEEoverridecommandlockouts 

\maketitle

\begin{abstract}
In recent years, interest in quantum computing has increased due to technological advances in quantum hardware and algorithms.
Despite the promises of quantum advantage, the applicability of quantum devices has been limited to few qubits on hardware that experiences decoherence due to noise.
One proposed method to get around this challenge is distributed quantum computing (DQC).
Like classical distributed computing, DQC aims at increasing compute power by spreading the compute processes across many devices, with the goal to minimize the noise and circuit depth required by quantum devices.
In this paper, we cover the fundamental concepts of DQC and provide insight into where the field of DQC stands with respect to the field of chemistry---a field which can potentially be used to demonstrate quantum advantage on noisy-intermediate scale quantum devices.

\end{abstract}


\section{Introduction}
Distributed systems play a crucial role for modern computing needs, such as for cloud and high-performance computing tasks. 
In general, a collection of autonomous computing elements, that appear to be a single system to the end-user, are called distributed systems.
Autonomous computing elements can range from two devices connected in a single room to hundreds or thousands of devices connected across vast spaces. 
When a distributed system behaves as expected, this is referred to as a coherent distributed system.
Within the realm of distributed systems, there are several types of high-performance distributed computing such as cluster, grid, and cloud computing~\cite{van_steen_distributed_2017}.

Chemistry, among other scientific fields, is one that has benefited by access to high-performance distributed computing.
Due to the many-body nature, of both classical simulations and electronic structure theory calculations, parallelism is required to solve chemical problems in a timely manner~\cite{lehtola_free_2022}.
Many aspects of electronic structure theory calculations can be accelerated by exploiting parallelism, either through traditional architectures based on central processing units (CPUs)~\cite{lotrich_parallel_2008}, acceleration hardware, like graphical processing units (GPUs)~\cite{gotz_chapter_2010}, or more recently, through cloud computing~\cite{raucci_interactive_2023}.
Quantum chemistry packages that are commonly used and applied on compute clusters, such as Gaussian~\cite{frisch_gaussian16_2016} and ORCA~\cite{neese_software_2022}, implement parallelism whenever possible. 
Certain packages, such as NWChem~\cite{apra_nwchem_2020} and LAMMPS~\cite{thompson_lammps_2022}, were developed at US national labs for large-scale calculations and simulations on national supercomputers, which are comprised of highly parallel nodes.

Within the classical regime, it was foreshadowed in 1929 by P. A. M. Dirac that ``\textit{the underlying physical laws necessary for the mathematical theory of a large part of physics and the whole of chemistry are thus completely known, and the difficulty is only that the exact application of these laws leads to equations much too complicated to be soluble}''~\cite{dirac_quantum_1929}.
While classical solutions for quantum chemistry rely on approximate schemes, in his now famous 1981 lecture ``\textit{Simulating Physics with Computers}'', Richard J. Feynman stated that ``\textit{nature isn't classical, dammit, and if you want to make a simulation of nature, you'd better make it quantum mechanical}''~\cite{feynman_simulating_1982}.
In this statement, Feynman alluded to using quantum computers as an alternative to classical computing for studying quantum mechanical systems. 
Over the last few decades, there has been a large push in the field of computational chemistry to map electronic structure problems onto quantum devices since quantum chemistry has been posed as a field where quantum advantage could be exhibited for classically intractable chemistry problems~\cite{aspuru-guzik_simulated_2005,cao_quantum_2019,mcardle_quantum_2020}.
To this end, the quantum phase estimation (QPE)~\cite{kitaev_quantum_1995} and variational quantum eigensolver (VQE)~\cite{peruzzo_variational_2014,bharti_noisy_2022} algorithms have been developed and applied to recover the energies and properties of atoms and molecules.

While quantum algorithms and hardware are rapidly improving, one emergent field, which chemistry may be well suited as a ``litmus test'' for, is the field of distributed quantum computing (DQC).
Within this field, monolithic code stacks, based on deep circuit depths with few qubits, could be moved to distributed code stacks running on many quantum processing units (QPUs) with shallow circuits, few qubits, and reduced noise to achieve computational speedups.
Herein, we will discuss the hardware and software implementations of DQC and the current and future applications of DQC for chemical applications.

\section{Distributed Quantum Computing}

\subsection{Theoretical Background}
Distributed quantum computing (DQC) offers an approach to increase the computational power of quantum computers by joining the computing power of multiple quantum processing units (QPUs), instead of one large device~\cite{barral_review_2024,denchev_distributed_2008}.
Often these tasks are performed between unknown quantum states at distant nodes in a quantum network.
Quantum networks, like their classical counterparts, coordinate and distribute information across devices.
One limitation of quantum computing, referred to as the \textit{no-cloning theorem}, is that arbitrary quantum states cannot be copied, i.e., quantum information cannot be broadcast or replicated.
Despite this limitation, there are ways to transmit and control quantum information reliably using entanglement-based distributed algorithms.
These algorithms offer unique and attractive approaches for addressing the limitations of classical computing, but these algorithms depend on the efficiency and reliability of the entangled states~\cite{denchev_distributed_2008,barral_review_2024}.

Quantum entanglement is a fragile correlation between a joint quantum system, stronger than any classical counterpart, and arises from various physical phenomena, ranging from the position and momentum of free particles to the energy levels of trapped ions.
As an example, consider the Bell states, also known as Einstein–Podolsky–Rosen (EPR) pairs, which are pure entangled quantum states
\begin{align}
    \ket{\Phi^{\pm}} &= \frac{1}{\sqrt{2}} \left( \ket{0}_{A} \ket{0}_{B} \pm \ket{1}_{A} \ket{1}_{B}  \right)\label{eq:DV_Bell states1} \\
    \ket{\Psi^{\pm}} &= \frac{1}{\sqrt{2}} \left( \ket{0}_{A} \ket{1}_{B} \pm \ket{1}_{A} \ket{0}_{B}  \right),
    \label{eq:DV_Bell states2}    
\end{align}
where two qubits, $\ket{0}$ and $\ket{1}$, are shared by two parties, Alice ($A$) and Bob ($B$).
These states are considered the simplest form of quantum entanglement between two qubits since the outcome of Alice's measurement determines the outcome of Bob's measurement~\cite{nielsen_quantum_2010}.

Next, we will discuss how quantum entanglement and classical communications can be used as a DQC resource.
Quantum teleportation uses entanglement to facilitate the distribution of quantum states to encode quantum information.
For DQC, two main variants exist: one-directional communication (quantum teleportation or teledata) and gate operations at a distance (gate teleportation or telegate)~\cite{caleffi_distributed_2022,barral_review_2024}.
Both teledata and telegate consume one EPR pair and the transmission of two classical bits.
For both methods, the performance depends on the remote operations performed on the quantum circuit, the network interconnection between the remote quantum processors, and the ratio of data and communication qubits.
For another remote operation to be performed, a new Bell state must be distributed among the remote processors.
The distributed communication overhead can be reduced by including more communication qubits, which allows more remote operations to be performed in parallel.
Subsequently, when more communication qubits are on a QPU, fewer data qubits are available.
Other costs include when the path between the remote operations of two processors is longer, the more Bell states that are consumed, and that entanglement swapping consumes the Bell state.
Thus, a trade-off arises with entanglement swapping between ``augmented connectivity'' and ``EPR costs''.
Within distributed Bell states, noise is distributed among the overall DQC.
This noise can be limited by methods known as entanglement distillation or purification\cite{hu_long-distance_2021,bennett_purification_1996}.

\begin{figure}[!t]
    \centering
    \includegraphics[width=\columnwidth]{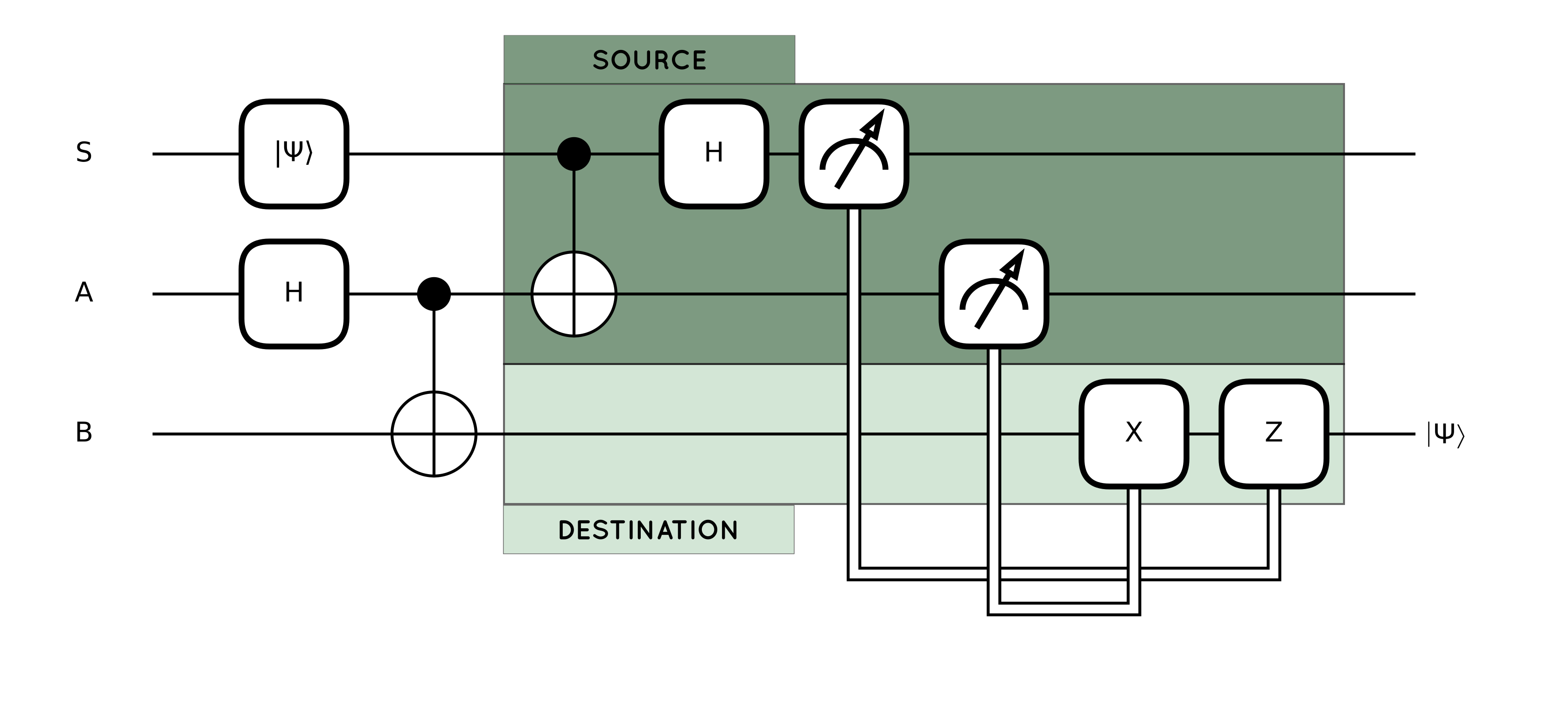}
\caption{An example of a quantum teleportation circuit, where the first two wires belong to the source and the bottom to the destination. \textit{Adapted from D. Cuomo et al. Towards a distributed quantum computing ecosystem. IET Quantum Communication, vol. 1, no. 1, pp. 3–8, 2020.}}
\label{fig:cuomo_towards_2019_fig6}
\end{figure}

The teledata communication paradigm generalizes the concepts of moving the states of qubits to remote devices.
Within this paradigm, qubits can be classified as either communication qubits, those used for generating entanglement, or data qubits, those used for data processing and storage.
It should be noted that an unknown qubit cannot be duplicated, observed, or measured without the qubit being altered.
Within the teledata scheme, a pair of parallel resources are required, where one source is classical, such that two bits of classical information can be transferred from source to destination. 
The other source must be quantum, containing an entangled pair of qubits shared between the source and destination nodes.
As an example, consider the teledata circuit shown in Fig.\ref{fig:cuomo_towards_2019_fig6}, where the top two wires belong to the source, with a generic state $\ket{\psi}$, and the bottom wire belongs to the destination, where after teleportation, $\ket{\psi}$ is available. 
And the EPR pair, $\ket{\Phi^{+}}$, consists of a couple of entangled qubits shared across the source and destination. 
The measurement is transmitted from the source to the destination using a classical link and the resulting classical bits are used for determining whether a bit- or a phase-flip, $X$ or $Z$, respectively, must be used to recover $\ket{\psi}$ from the EPR pair member available at the destination~\cite{cuomo_towards_2020}.

\begin{figure}[!t]
    \centering
    \includegraphics[width=\columnwidth]{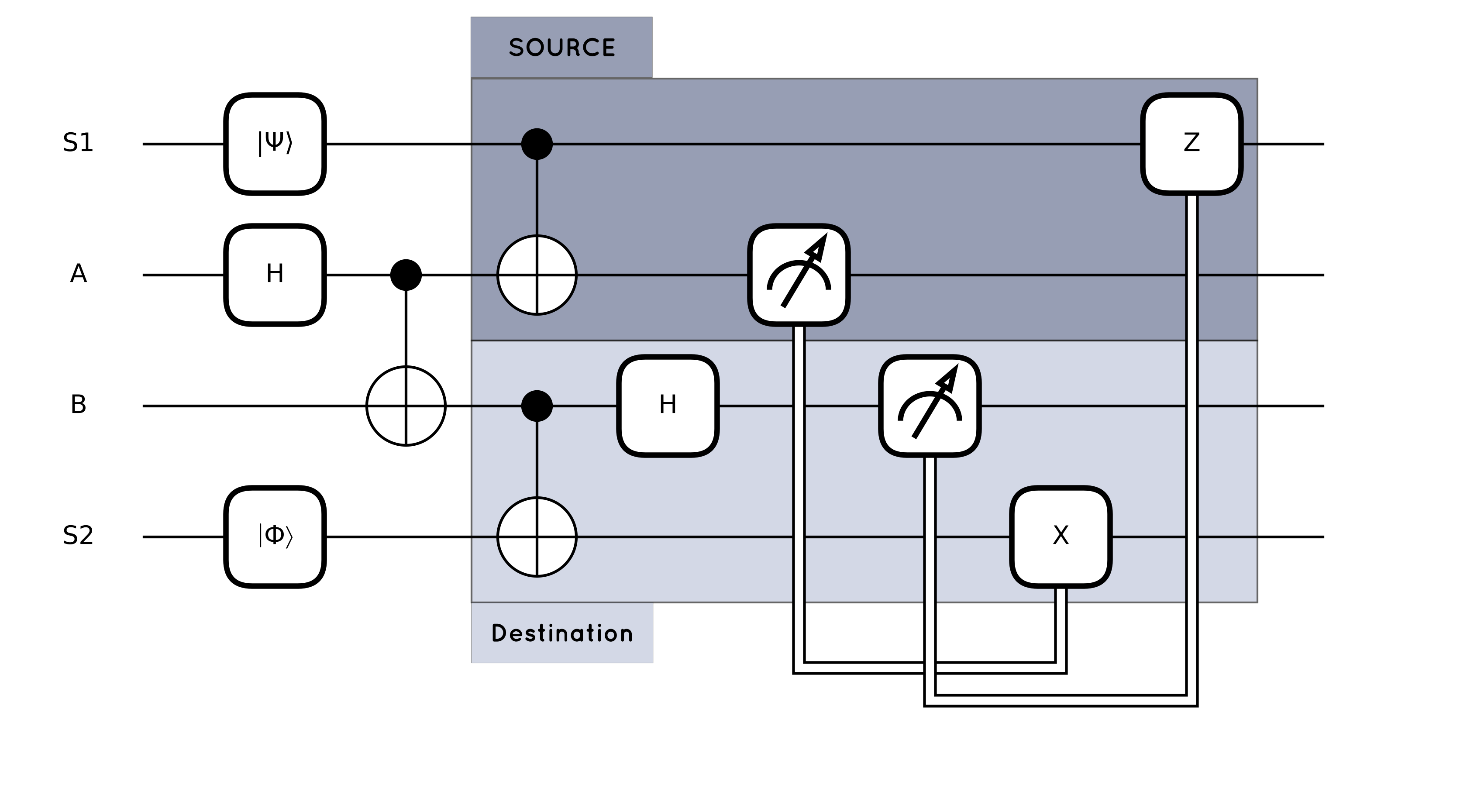}
    \caption{An example of a quantum telegate circuit that implements a $CNOT$ operation between remote qubits. \textit{Adapted from D. Cuomo et al. Towards a distributed quantum computing ecosystem. IET Quantum Communication, vol. 1, no. 1, pp. 3–8, 2020.}}
    \label{fig:cuomo_towards_2019_fig8}
\end{figure}

The quantum gate teleportation, or telegate, paradigm replaces two-qubit gates with auxiliary entangled states, local measurements, and single-qubit operations to reduce the resource requirements.
Unlike teledata, which transports quantum information from one device to another, through an entangled pair, telegate works by teleporting gates from device to device.
An example of a telegate circuit where $CNOT$ operations are shared between remote qubits is shown in Fig. \ref{fig:cuomo_towards_2019_fig8}.
Generic initial states, $\ket{\psi}$ and $\ket{\phi}$, are initially stored on the first and fourth qubits, respectively, whereas the EPR pair, $\ket{\Phi^{+}}$, are stored on communications qubits.
The $CNOT$ operations are performed between qubits on different devices, i.e., the first (source) and fourth (destination) qubits, shared between the EPR pair stored on the same device.
Lastly, it should be mentioned that a generalization of quantum teleportation using multipartite entangled states, or Greenberger-Horne-Zeilinger (GHZ) states, can be used for assisted and unassisted $N$-party teleportation~\cite{cuomo_towards_2020,barral_review_2024}.


\subsection{Physical Hardware and Hardware Distribution Layers}
\begin{figure}[!t]
\centering
\includegraphics[width=\linewidth]{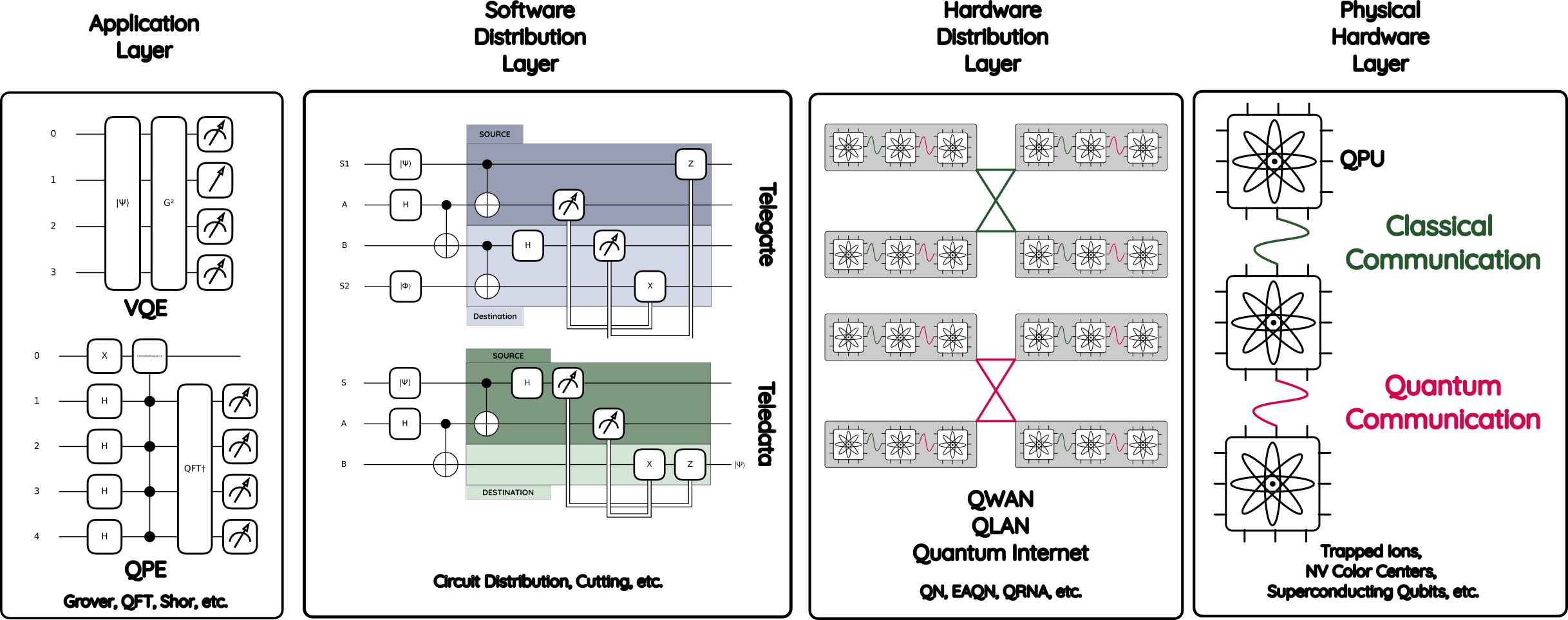}
\caption{An example of a layered model of distributed quantum computing (DQC) which includes the physical hardware, hardware distribution, software distribution, and application layers. \textit{Inspired by D. Barral et al. Review of Distributed Quantum Computing. From Single QPU to High-Performance Quantum Computing. arXiv:2404.01265, 2024.}}
\label{fig:Images_chapter1_abstract_figure_3_with_qpus}
\end{figure}

Concerning the physical hardware layer (Fig. \ref{fig:Images_chapter1_abstract_figure_3_with_qpus} far right layer), there are many diverse options for QPUs, such as those based on superconducting qubits with short gate operations and NV color centers in diamond with long qubit coherence~\cite{krantz_quantum_2019,pezzagna_quantum_2021}.
Since the homogenization of quantum computing platforms is unlikely, specialized, modular architectures have been proposed to maximize the performance of DQC~\cite{barral_review_2024}.
Within this layer, there are five types of devices: QPUs, quantum transducers, quantum memories, quantum repeaters, and entanglement routers and switches~\cite{caleffi_distributed_2022}. 
Fundamentally, QPUs are intended to be integrated into high-performance computing (HPC) infrastructures, like CPUs or GPUs. 
QPUs refer to a single device, where the quantum algorithms are performed by qubit operations~\cite{saurabh_conceptual_2023,wintersperger_qpu-system_2022,vazquez-perez_qpu_2024,mccaskey_extending_2021,mccaskey_xacc_2019}.
Within a shared development environment, this integration allows for what is called quantum-centric supercomputing centers~\cite{noauthor_quantum-centric_nodate}.
Within the field of DQC, there are two types of systems: those with entanglement between nodes and those with classical inter-node communication across a full set of qubits~\cite{barral_review_2024}.
The physical hardware and hardware distribution layers, shown in Fig. \ref{fig:Images_chapter1_abstract_figure_3_with_qpus}, are equivalent to the lowest three (physical, data link, and network) layers of the classical Open Systems Interconnection (OSI) model~\cite{1400-1700_isoiec_nodate}.
In this context, the mechanisms that allow for two physically separated QPUs to be connected are referred to as the physical hardware layer and the hardware distribution layer allows for communication between multiple QPUs~\cite{barral_review_2024}.
The communication between QPUs can be handled according to a coupling map to help mitigate decoherence and noise~\cite{cuomo_towards_2020}.
To allow for communication between local qubit devices, such as QPUs, memories, or repeaters, quantum transducers take the quantum information and transform it into a transmittable signal~\cite{grobe_wavelength_2013,zhong_optically_2015,edlbauer_semiconductor-based_2022}.
To store quantum information over a long period, long-lived qubits, such as trapped ions, can be used for quantum memories or quantum Random Access Memory (qRAM)~\cite{heshami_quantum_2016}.
Another aspect that allows entanglement operations at a distance are quantum repeaters~\cite{azuma_quantum_2023}. 
These devices must be reliable to perform deterministic teleportation and distribution protocols of quantum states and remote operations between nodes.
Lastly, as we move towards the hardware distribution layers of DQC, entanglement routers and switches allow teleportation between arbitrary parts of the distributed system by establishing entanglement between compute nodes~\cite{lee_quantum_2022}.

For the hardware distribution layer, a quantum local area network (QLAN) can be used to connect one or more switches and/or routers, while a quantum wide area network (QWAN) can be used to route between multiple routers.
The creation and distribution of entanglement between QPUs, whether within closer proximity or long distances, can be handled by quantum networks (QNs).
One such example of a global QN would be the Quantum Internet, which would enable communication between remote quantum nodes and could provide a fundamental communication infrastructure for DQC~\cite{gyongyosi_advances_2022,renger_cryogenic_2023}.
Using a distributed approach, through the Quantum Internet, would allow the realization of large-scale quantum processors that mimic modern HPC environments based on thousands of inter-connected processors, memories, and storage units.
It has been proposed that with a linear amount of physical resources, using interconnected QPUs, the Quantum Internet could achieve exponential speed-ups~\cite{cuomo_towards_2020}.

\subsection{Software Distribution and Application Layers}
The next layers, the application and software distribution layers (Fig. \ref{fig:Images_chapter1_abstract_figure_3_with_qpus} left two layers), are the most relevant for chemical applications.
Like its classical counterpart, quantum compilation is a key aspect of the software distribution layer, which involves optimizing compute resources to produce efficient code and translating programs into machine-readable instructions.
Quantum compilation is largely a classical task, where quantum development software is built on top of existing classical languages.
For distributed computing, compilation comes with increased overhead related to communications and synchronization between processing units.
For DQC, there are three types of non-mutually exclusive, distribution:  circuit distribution~\cite{kernighan_efficient_1970,akhremtsev_engineering_2017,daei_optimized_2020,schlag_high-quality_2023}, circuit cutting~\cite{lowe_fast_2023,peng_simulating_2020}, and embarrassing parallelism~
\cite{peruzzo_variational_2014,schuld_evaluating_2019,viqueira_density_2023,wierichs_general_2022,stokes_quantum_2020}.
Circuit distribution is associated with a quantum communication network that permits the execution of a circuit that demands more qubits than is available on a single QPU.
This involves finding an optimal partition, distributing the partition among available QPUs, and mapping the partition to each QPU.
While it has been proposed that quantum communication within a quantum network is required for DQC, circuit cutting has been proposed as a solution for partitioning a circuit that is too large to fit on a single QPU into smaller units with no entanglement, i.e., quantum communication is unavailable.
These subcircuits can be simulated classically, either sequentially or in parallel, and the output of the original circuit can be recovered using the results of the subcircuits.
Lastly, an application may be embarrassingly parallel, if no quantum communication exists and the circuit fits on one QPU.
In this case, a problem is divided into smaller, independently executed computations without the need for direct communication between them~\cite{barral_review_2024}.
It should also be noted that, a combination of distribution types has been attempted through the Quantum Divide and Conquer Algorithm (QDCA) proposed by Tomesh \textit{et al.}~\cite{tomesh_divide_2023}.

\section{Chemical Applications and Outlook}

Lastly, we will discuss the current applications of DQC in chemistry and possible future developments in the field.
Most of the distributed quantum algorithms found in the literature focus on the application of the Simon~\cite{tan_distributed_2022,avron_quantum_2021}, Deutsch-Jozsa~\cite{avron_quantum_2021,li_distributed_2024}, Grover~\cite{zhou_distributed_2023,srivastava_using_2024}, Bernstein-Vazirani~\cite{zhou_distributed_2023-1}, Shor~\cite{yimsiriwattana_distributed_2004}, and leader election~\cite{denchev_distributed_2008} algorithms.
While the previously mentioned algorithms do not have any relevance to the field of chemistry, other examples, based on quantum phase estimation (QPE), could be relevant for chemical applications, since the QPE algorithm can be used to determine the energy of a trial wave function~\cite{aspuru-guzik_simulated_2005,cirac_distributed_1999,neumann_imperfect_2020,shi_reference_2023}.
On current devices, the variational quantum eigensolver (VQE) algorithm, is a classical-quantum hybrid algorithm, which is more attractive for recovering the ground state energies of molecules~\cite{cao_quantum_2019}.
As previously mentioned, circuit cutting can be used as a platform for distributed computing, and in combination with VQE can be used to simulate the ground state energy of BeH$_{2}$~\cite{peng_simulating_2020}.
Another example, based on entanglement forging to double the amount of available qubits, was used to accurately simulate the ground state potential energy surfaces of H$_{2}$O~\cite{eddins_doubling_2022}.

One of the most promising examples of DQC within the field of chemistry is based off of a quantum-centric supercomputing architecture~\cite{alexeev_quantum-centric_2024}.
This architecture uses 6400 nodes on the Fugaku supercomputer, in tandem with the IBM superconducting Heron quantum processor.
Within the Heron processor, 58, 45, and 77 qubits were used to simulate the bond breaking of N$_{2}$ and active space calculations for clusters such as, [Fe$_{2}$S$_{2}$(SCH$_{3}$)$_{4}$]$^{-2}$ and [Fe$_{4}$S$_{4}$(SCH$_{3}$)$_{4}$]$^{-2}$, respectively.
The maximum circuit depth is reported to be 10,570, with 3,590 of them corresponding to 2-qubit quantum gates.
A highlight of this work is that the intrinsically quantum components of the calculations are outsourced to the QPU, while classical distributed computing is allowed to reduce the workflow on the QPU~\cite{robledo-moreno_chemistry_2024}.
This type of distributed scheme can provide a path toward quantum advantage by off-loading portions of an algorithm for which quantum hardware is particularly suited for, while allowing HPC to handle the classical portions of the algorithm efficiently.

While tremendous progress has been made in recent years concerning novel theory advances for quantum computing~\cite{grimsley_adaptive_2019,stair_multireference_2020,feniou_overlap-adapt-vqe_2023}, quantum chemistry is largely still held to the field of classical computing~\cite{cao_quantum_2019,mcardle_quantum_2020}.
Within this field, novel parallelization algorithms, acceleration hardware, and access to HPC architectures have allowed classical electronic structure theory to proliferate.
Common quantum distribution tasks such as the distribution of observable terms, gradient and Hessian terms, the data, and gradient-free optimizations could easily be adapted to subroutines that are commonly found in quantum chemistry packages related to configuration interaction-based problems~\cite{vogiatzis_pushing_2017,gao_distributed_2024}, integral evaluation~\cite{luehr_dynamic_2011,yasuda_two-electron_2008}, and algorithms for the recovery of variational energies~\cite{pulay_improved_1982} and geometry optimizations~\cite{csaszar_geometry_1984}~\cite{barral_review_2024}.
Additionally, due to the increased application of machine learning in chemistry, algorithms for distributed quantum machine learning could also be applied to chemical applications~\cite{marshall_high_2023,barral_review_2024}.


While there have been many advancements related to both hardware and software aspects of DQC, many challenges remain before an advantage can be shown for chemical applications.
Most of the challenges, concerning hardware, are the same challenges that noisy intermediate-scale quantum (NISQ) devices face, such as decoherence and noise.
The challenges for software development are related to the immediate application of distributed algorithms and compilers to chemical problems
Within this field, the implementation of distributed algorithms has been limited, with the previously mentioned study~\cite{robledo-moreno_chemistry_2024} being among the first papers to apply DQC for chemical applications.
Integrating existing distributed quantum algorithms into chemists' existing workflows, who perform electronic structure calculations or quantum machine learning, could show potential advantages for the field.
Another challenge would be moving quantum chemistry packages from software created for classical distributed systems to DQC, or those currently interfaced with quantum computing packages but not yet applied to DQC.
While these challenges can be daunting, they can be addressed through future collaborations, where computational chemists can work in tandem with experts on distributed systems and QC to integrate their desired algorithms on DQC, and through group initiatives to push the field beyond single QPUs to quantum-centric supercomputing.




\bibliography{quantum_week}

\end{document}